\documentclass[11pt,twoside]{article}
\usepackage{asp2010}

\resetcounters

\begin{document}

\title{The hunt for old novae}
\author{C. Tappert,$^1$ N. Vogt,$^1$ L. Schmidtobreick$^2$, and 
A. Ederoclite$^3$
\affil{$^1$Dpto. de F\'{\i}sica y Astronom\'{\i}a, Universidad de 
Valpara\'{\i}so, Avda. Gran Breta\~na 1111, Valpara\'{\i}so, Chile}
\affil{$^2$European Southern Observatory, Alonso de Cordova 3106, Santiago, 
Chile}
\affil{$^3$Centro de Estudios de F\'{\i}sica del Cosmos de Arag\'on, Plaza San 
Juan 1, Planta 2, Teruel, E44001, Spain}}

\begin{abstract}
We inform on the progress of our on-going project to identify
and classify old classical novae, using deep {\em UBVR} photometry and
subsequent spectroscopy for a proper candidate confirmation, and
time-resolved observations in order to find the orbital period and other
physical properties of the identified old novae. This way, we have already
increased the number of confirmed southern and equatorial post-novae from
33 to 50, and determined new orbital periods of eight objects. As an example,
we summarise the results on V728 Sco (Nova Sco 1862) which we found to be
an eclipsing system with a 3.32 h orbital period, displaying high and low
states similar to dwarf-nova outbursts. Analysis of the low-state eclipse 
indicates the presence of a small hot inner disc around the white dwarf 
component.
\end{abstract}

\section{Tracking down the CV within the nova}

\subsection{The species}
It is undisputed that a nova eruption represents an event in a cataclysmic
variable (CV). The old nova DQ Her (Nova Her 1934) has become known as the
prototype intermediate polar \citep{warner83-1}, RR Pic (Nova Pic 1925) shows
all the properties of an SW Sex star \citep{schmidtobreicketal03-4}, and
there is compelling evidence for ancient nova shells around the dwarf novae
Z Cam and AT Cnc \citep{sharaetal07-1,sharaetal12-4}. In principle, every
CV should experience nova eruptions, as long as its mass-transfer rate is
sufficiently high to accumulate the necessary mass on the white dwarf during
its lifetime. 

Still, there are several unresolved questions concerning the relation between
novae and CVs. Which CV parameters are favourable for the system to undergo
a nova eruption? The mass-transfer rate is easily identified as an important
one, but what about, e.g., the strength of the magnetic field of the white
dwarf? Another point concerns the long-term consequences of the nova eruption.
Do the white dwarfs gain or lose mass in the long run? And how does the
eruption affect the mass-transfer rate, i.e. do post-novae go into hibernation
\citep{sharaetal86-1,prialnik+shara86-1}?

One way to find an answer is to study the post-novae as a group and to
compare with the general CV population. However, this needs samples of 
a statistically significant size to distinguish the rule from the exception,
and the novae are severely lacking in this respect. In the following we
restrict our research to novae that have erupted before 1980, in order to 
allow $\sim$30 yr for the contribution of the ejected material to become 
negligible (at least in the optical range), so that the properties of the 
underlying CV can be studied. The catalogue of CVs by \citet{downesetal05-1} 
lists 200 such systems, but only 28 of these have good orbital
periods\footnote{This includes the period of CP Pup, which is one of the
few below the period gap, and whose orbital nature is now reported to be
doubtful; see the contribution by Mason et al. in these proceedings.},
and about 3/4 of the reported novae still lack spectroscopic confirmation or
even the identification of a candidate for the post-nova.

\subsection{Picking up the scent}
\begin{figure}
\centerline{\includegraphics[width=0.9\columnwidth]{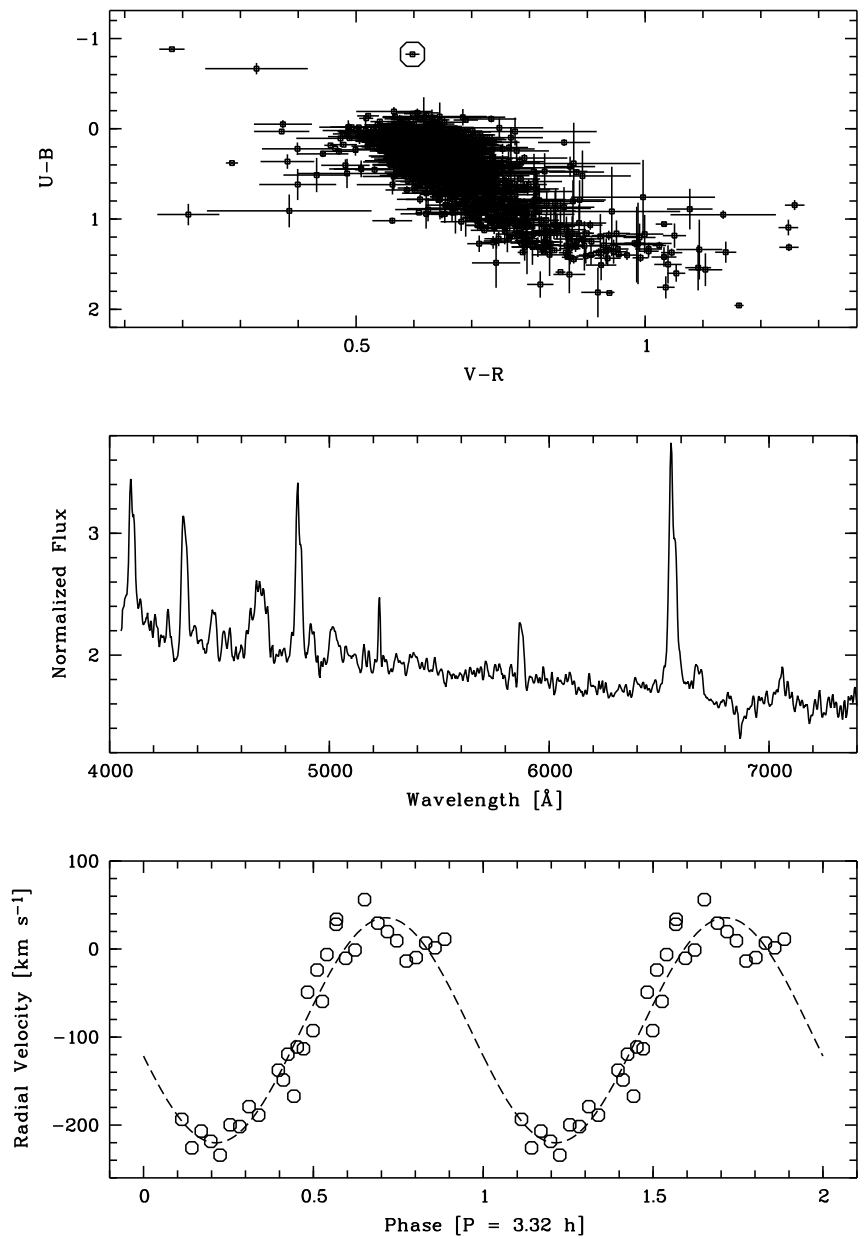}}
\caption[]{The old nova V728 Sco as an example. 
{\em Top:} {\em UBVR} photometry of a 4.5$^\prime$ $\times$ 4.5$^\prime$ 
field centred on the reported coordinates. The nova (marked by a circle)
turned out to be $\sim$2$^\prime$ NW of that position. 
{\em Middle:} Low-resolution spectroscopy of the identified candidate. 
The spectrum shows prominent Balmer, He{\sc i} and He{\sc ii} emission lines.
{\em Bottom:} Radial velocities of the H$\alpha$ emission line folded on the 
orbital period ($P_\mathrm{orb} = 3.32~\mathrm{h}$). The dashed curve
represents the best sine fit to the data.
\label{v728sco1_fig}
}
\end{figure}

In order to improve this situation we have started a program to identify 
nova candidates, to confirm them spectroscopically and to determine their
orbital period. First results and a detailed description can be found in
\citet{tappertetal12-1}. For the identification of nova candidates we
take advantage of CVs being three-component systems (white dwarf, late-type
star, accretion disc or stream), placing them apart from the main-sequence
in a colour-colour diagram. We therefore use {\em UBVR} photometry to
select the candidates, low-resolution spectroscopy to examine them for the
presence of typical emission lines, and time-series spectroscopy and/or
photometry to measure the orbital period. As an example for a system that
went through all our different bins (unidentified candidates -- unconfirmed
candidates -- confirmed post-novae without orbital period -- systems with
an established orbital period) we present the data on V728 Sco in 
Fig.~\ref{v728sco1_fig}. We will come back to that object in Section 
\ref{v728sco_sec}.

\subsection{Game count then and now}
At the start of our program in 2009, the 153 reported pre-1980 novae on 
the southern hemisphere ($\delta \le +20^\circ$) included 120 unconfirmed
objects, 86 of them without a proper candidate. For 24 post-novae a value
for the orbital period was listed. Almost four years and several observing
runs (not much favoured by meteorological conditions) later, we have increased 
the number of confirmed post-novae to 50 (from 33) and determined the orbital
period for eight objects. One reported nova, V734 Sco, turned out to be
a Mira variable whose photometric variation was mistaken for a nova eruption%
\footnote{In \citet{tappertetal12-1} we had reported the same conclusion for
V1310 Sgr, but that candidate is probably not identical to the reported nova 
(Schaefer, private communication).}. This 
still leaves about 2/3 of the sample of pre-1980 novae as unconfirmed objects, 
and further observations are in preparation to significantly decrease that 
number.

\section{V728 Sco: a rare animal?}
\label{v728sco_sec}

\begin{figure}
\centerline{\includegraphics[width=0.9\columnwidth]{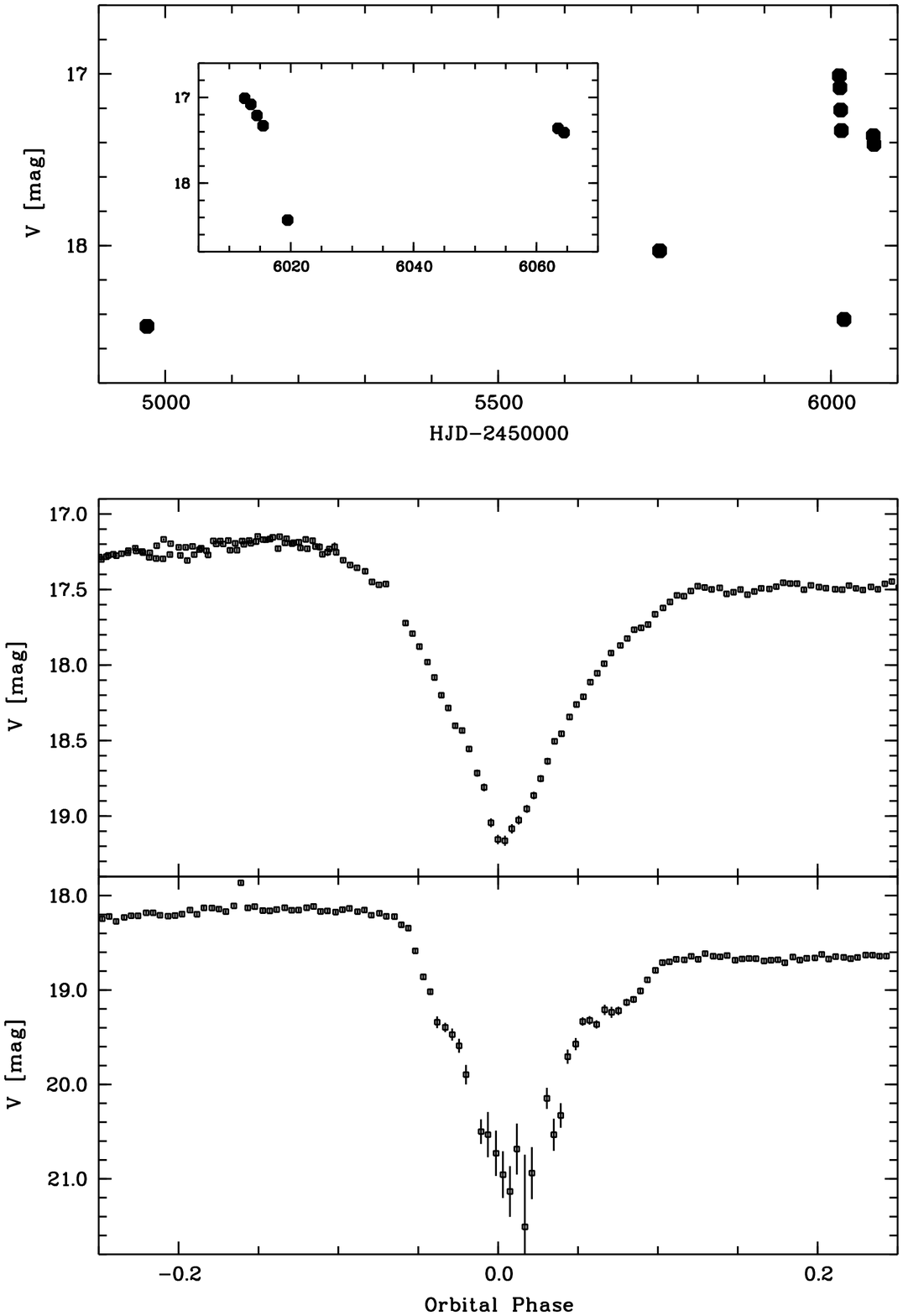}}
\caption[]{Time-series photometric data on V728 Sco. 
{\em Top:} Long-term light curve (out-of-eclipse data) from 2009 May 20 to 
2012 May 16. The inset presents a close-up of the 2012 March to May data.  
{\em Middle:} High-state eclipse from 2012 March 28.
{\em Bottom:} Low-state eclipse from 2012 April 1. Note the different y-scale
in the eclipse data.
\label{v728sco2_fig}
}
\end{figure}

The system V728 Sco corresponds to a nova eruption from 1862
\citep{tebbutt78-3}. This makes it the third oldest confirmed post-nova
(not counting the likes of Z Cam and AT Cnc) and thus an interesting
object to examine the consequences of the nova eruption on the underlying
CV. A detailed study can be found in \citet{tappertetal13-1}. Here we present
a summary of the main results.

Most known post-novae are high mass-transfer systems, with their spectra
being characterised by a steep blue continuum and weak emission lines
\citep[e.g.,][]{ringwaldetal96-3}. The spectrum of V728 Sco, however,
is more reminiscent of that of a dwarf nova: a moderately blue continuum and 
strong emission lines of the Balmer and He{\sc i} series (middle plot in
Fig.~\ref{v728sco1_fig}). Such properties let suspect a comparatively
low mass-transfer rate. The only feature that appears out of place for a
dwarf nova is the strong Bowen/He{\sc ii} blend at $\lambda$4641--4686 {\AA},
which indicates the presence of a hot component somewhere in the system.
The detection of long-term variations that resemble dwarf-nova outbursts
(top plot in Fig.~\ref{v728sco2_fig}) represents further evidence for
a comparatively low mass-transfer rate. 

Time-series data show the object to be eclipsing with an orbital period
$P_\mathrm{orb} = 3.32~\mathrm{h}$. This places V728 Sco within the
regime of the SW Sex stars, high mass-transfer systems that dominate the
orbital period range 2.8--4 h \citep[][Schmidtobreick et al., these 
proceedings]{rodriguez-giletal07-1,rodriguez-giletal07-2}. Thus the more 
puzzling is the low mass-transfer behaviour of V728 Sco.

In high state, the triangular shape of the eclipse indeed is very similar
to that of SW Sex stars \citep[e.g.,][]{stanishevetal04-2}. However,
in low state the eclipse is much more revealing, showing the ingress and
egress of the hot spot and of the central object (Fig.~\ref{v728sco2_fig}). 
While the latter usually is expected to be the white dwarf, our analysis of 
the ingress and egress times yields a radius for that central component of 
$R \sim 0.09~R_\odot$, significantly larger than a white dwarf. We conclude 
that this component represents a hot and optically thick inner disc, that
is caused by irradiation from the still eruption-heated white dwarf. The
presence of such disc would also explain the small-amplitude and high-frequency
nature of the outbursts \citep{schreiberetal00-2}.

While most old novae indeed seem to be high mass-transfer systems
\citep{ibenetal92-1}, there are some known exceptions, e.g. the old novae
XX Tau \citep{schmidtobreicketal05-2} and V446 Her \citep{honeycuttetal95-1}.
Due to its eclipsing nature V728 Sco certainly appears as the member of this
group that is the most interesting for further studies. It is most desirable
to monitor the long-term behaviour more extensively to examine the frequency
and the stability of the outbursts. \citet{schreiberetal00-2} predict 
increasing amplitude and decreasing frequency over time, as the white 
dwarf continues to cool down. Additionally, high S/N and high time-resolved
data (especially in low state) should be able to determine accurate 
physical system parameters of V728 Sco.

\acknowledgements This research was supported by FONDECYT Regular grant 
1120338 (CT and NV). AE acknowledges support by the Spanish Plan Nacional de 
Astrononom\'{\i}a y Astrof\'{\i}sica under grant AYA2011-29517-C03-01.
We gratefully acknowledge ample use of the SIMBAD database, 
operated at CDS, Strasbourg, France, and of NASA's Astrophysics Data System 
Bibliographic Services. The observational data for this program were taken with
ESO telescopes in La Silla and Paranal. Thanks to Christian Knigge for pointing
out the similarity between V728 Sco's high-state eclipse light curve 
and that of SW Sex stars, and to Brad Schaefer for the information on V1310 
Sgr.


\end{document}